\documentclass{aa}
\usepackage{graphicx}
\usepackage{epsfig}
\newcommand{\msun}{\mbox{$M_{\odot}$}}
\newcommand{\Msun}{\mbox{$M_{\odot}$}}
\newcommand{\lsun}{\mbox{$L_{\odot}$}}
\newcommand{\Lsun}{\mbox{$L_{\odot}$}}

\newcommand{\zsun}{\mbox{$Z_{\odot}$}}
\newcommand{\teff}{\mbox{$T_{\rm eff}$}}
\newcommand{\Teff}{\mbox{$T_{\rm eff}$}}

\newcommand{\vinf}{\mbox{$v_{\infty}$}}
\newcommand{\vesc}{\mbox{$v_{\rm esc}$}}
\newcommand{\ratio}{\mbox{$v_{\infty}$/$v_{\rm esc}$}}
\newcommand{\mdot}{\mbox{$\dot{M}$}}
\newcommand{\Mdot}{\mbox{$\dot{M}$}}

\newcommand{\msunyr}{\mbox{$M_{\odot} {\rm yr}^{-1}$}}
\newcommand{\kms}{km~${\rm s}^{-1}$}

\newcommand{\bb}{\bibitem[]{}}
\begin{document}


\title{Hot Horizontal Branch Stars: Predictions for Mass Loss}

\subtitle{Winds, rotation, and the low gravity problem}
\author{Jorick S. Vink\inst{1,2} \and Santi Cassisi\inst{3}}
\offprints{Jorick S. Vink, jvink@ic.ac.uk}

\institute{Imperial College, Blackett Laboratory, 
           Prince Consort Road, London, SW7 2BZ, U.K.
           \and 
           Astronomical Institute, Utrecht University,
           P.O.Box 80000, NL-3508 TA Utrecht, The Netherlands
	   \and
           I.N.A.F. - Osservatorio Astronomico di Collurania, 
           Via M. Maggini, 64100 Teramo, Italy}

\date{Received 21 May 2002; accepted 17 June 2002}

\titlerunning{Hot HB Stars: Predictions for Mass Loss}
\authorrunning{Jorick S. Vink \& Santi Cassisi}

\abstract{We predict mass-loss rates 
for the late evolutionary phases of low-mass stars,  
with special emphasis on the consequences for the morphology of 
the Horizontal Branch (HB).
We show that the computed rates, as predicted by the most plausible mechanism 
of radiation pressure on spectral lines, are too low to produce EHB/sdB 
stars. This invalidates the scenario recently outlined by Yong et al. (2000) 
to create these objects by mass loss {\it on} the HB. 
We argue, however, that mass loss plays a role in the distribution 
of rotational velocities of hot HB stars, and may  
-- together with the enhancement of heavy element abundances due 
to radiative levitation -- provide an explanation for the so-called ``low gravity'' problem. 
The mass loss recipe derived for hot HB (and extreme HB, sdB, sdOB) stars 
may also be applied to post-HB (AGB-manqu{\'e}, UV-bright) stars over a range 
in effective temperatures between 12\,500 -- 40\,000 K. 
}

\maketitle


\section{Introduction}
\label{s_intro}

Over the last decades, both observational and 
theoretical efforts have been devoted to the investigation of  
the observed distribution of stars along the Horizontal 
Branch (HB) of galactic Globular Clusters (GCs). 
Although canonical stellar evolution theory has provided a 
general consensus on the evolutionary phase 
corresponding to the HB sequence, and convincingly 
demonstrated that its morphology is most strongly 
affected by cluster metallicity (the \lq{first}\rq\ parameter; 
Sandage \& Wallerstein 1960), 
many problems remain. 
The most striking controversy involves the wide variety 
of HB morphologies among clusters with similar 
metallicities (the \lq{second parameter}\rq~problem; Sandage \& Wildey 1967, 
van den Bergh 1967). 
Candidate second parameters are cluster age (e.g. Lee et al. 1994
and references therein), mass loss along the Red Giant Branch (RGB) 
(Catelan et al. 2001 and references therein), rotation and deep
helium mixing (Sweigart 1997), dynamical interactions 
involving binaries and even planets (Soker 1998), 
as well as environmental effects in high-density environments 
(Fusi Pecci et al. 1993).

The identification of the second parameter is especially relevant
to the formation of Extremely blue HB (EHB) stars, which 
are thought to be responsible for the ultraviolet upturn phenomenon
in elliptical galaxies (Greggio \& Renzini 1990; Dorman et al. 1995). 
The presence of EHB stars as blue \lq{tails}\rq~in clusters 
(Ferraro et al. 1998; Piotto et al. 1999), 
as well as sdB/sdO stars in the field (Greenstein 1971; 
Green et al. 1986), has inspired 
modern-day research to explain their formation both through 
mechanisms that produce high mass loss along the RGB (Soker et al. 2001 and
references therein), 
as well as through binarity (Mengel et al. 1976, Heber et al. 2002). 

Further puzzles in HB morphology concern the issues of 
HB \lq{gaps}\rq~(Newell 1973) -- specific regions 
along the branch that are significantly underpopulated\footnote{It has been 
claimed that the positions of the gaps along the HB in different galactic 
GCs are the same within current empirical uncertainties (Ferraro et al. 1998),
however it is not clear whether these gaps mark regions with 
specific effective temperatures, or whether they correspond to 
constant mass loci (Piotto et al. 1999). Note that Catelan et 
al. (1998) have challenged the existence of gaps at the same positions in 
all clusters.}, and a relatively new, unexplained, 
but ubiquitous feature is the so-called 
Str{\"o}mgren u-jump at an effective temperature of 
\teff\ $\simeq$\,11\,000 K (Grundahl et al. 1999), possibly coinciding
with a jump in $\log g$ (Moehler et al. 1995), and an unexplained
absence of fast rotators above this temperature (Behr et al. 2000; 
Recio-Blanco et al. 2002).  

As discussed by Grundahl et al. (1999), the Str{\"o}mgren u-jump may be due to 
atmospheric diffusion by radiative levitation of heavy 
elements, as both Glaspey et al. (1989) and Behr et al. (1999) 
found striking abundance anomalies in blue HB stars, with 
iron enhancements of up to three times the solar value.
Moehler et al. (2000) have shown that the enhancement of heavy elements 
in spectroscopic analyses may partially solve the problem of the 
anomalously low gravities along the blue HB, but the discrepancy 
is still present at the level of $\Delta\log{g}\,\approx\,0.1$ dex 
for stars in the range $15\,000\,<\teff\,<\,20\,000\,$K. 
Even more so, the first two mentioned HB features -- the blue 
tails and the gaps -- are still an enigma\footnote{Note that
Brown et al. (2001) and Sweigart et al. (2002) have provided a 
theoretical framework which could explain the hot gap in the HB of 
NGC\,2808.}, and it is not at all 
obvious whether they originate from a mechanism working in a prior 
evolutionary phase (on the RGB) or if they are due to a process 
working \lq{in situ}\rq\, once the star has settled on the HB.
One of the options that may help in explaining the above-mentioned 
problems is mass loss {\it on} the HB. 

It is worth mentioning that the colour width of 
the hottest gap is so small that changes of the order 
of a few times $10^{-3}$\,\Msun\ in the total mass are capable to 
move the star away from its initial location, 
far enough as to produce an underpopulated region 
in the H-R diagram (HRD). 
For this to occur one needs to identify a mass-loss 
mechanism which efficiency rapidly increases at the specific
effective temperatures of the gap.

The hypothesis that a mass-loss mechanism may be at work 
during the HB evolution was first entertained by 
Wilson \& Bowen (1984). They suggested that an increased 
mass-loss efficiency, when crossing the RR Lyrae instability strip, 
could provide an explanation for the HB mass distribution in a more 
natural way than the alternative of a stochastic variation 
in the amount of mass lost during the prior RGB phase. The topic of mass 
loss on the HB was further addressed by Koopmann et al. (1994), 
but they concluded that constant mass loss in the RR Lyrae strip 
was incapable of providing an explanation for the HB mass dispersion 
or the RR Lyrae period change distribution.
Additionally, mass loss during the central He-burning phase was suggested
by Michaud et al. (1985) and Bergeron et al. (1988) in order to explain 
the large silicon underabundances in some HB stars. 
More recently, Yong et al. (2000) performed accurate evolutionary computations 
with mass loss, and suggested that mass-loss rates of the order 
of $10^{-9}$ -- $10^{-10}$ \msunyr\ for HB stars in the metal-rich 
cluster NGC\,6791 can force these stars to move to a bluer position 
and thus lead to the production of EHB stars. 
If correct, this scenario could provide an explanation for the
presence of extended blue tails along the HB of some metal-rich GCs, such 
as NGC\,6441 and NGC\,6388 (Rich et al. 1997), although it would not 
be able to explain the upward sloping of the HB in these clusters 
(Raimondo et al. 2002 and references therein).
The main problem with the proposed scenario, however, 
is that no physical mechanism for mass loss was proposed and 
that the adopted mass-loss rates were completely \lq{ad hoc}\rq, as there 
are neither observational data indicative of mass loss on the HB available, 
nor any predictions. 

Our aim in the present paper is to alleviate current shortcomings by 
computing radiation-driven wind models and mass-loss rates for low-mass blue 
stars, and to subsequently investigate their influence on 
HB evolutionary models. Blue HB stars are located in a region of 
the HRD, where the stars are hot (with \teff\ between 10\,000 and 35\,000 K), 
and relatively bright, and radiation pressure 
forces can therefore be considered a natural driving mechanism. 
Although there may be other processes that could possibly drive a wind, 
such as pulsations 
\footnote{Recently a new class of variable stars has been discovered 
in the field, the so-called EC14026 (Kilkenny et al. 1997), which have been identified 
as hot HB stars and their progeny. Even though no clear identification 
of similar variables in GCs have been obtained, there is 
no {\it a priori} reason for the lack of this kind of pulsation among 
cluster HB stars.}, 
all other wind-driving options are much less well-understood than 
radiation pressure on spectral lines. 
Radiation-driven wind models have been developed in the 1970s 
by Lucy \& Solomon (1970) and Castor et al. (1975). In more recent 
days, the models have been very successful in predicting the values
observed in O supergiants (Vink et al. 2000). The direct application of these 
predictions to HB stars, such as the use of the mass-loss recipe provided by Vink 
et al. (2000) would however involve a rather large and dangerous extrapolation 
by four orders of magnitude in stellar luminosity. 

As far as the \lq{gaps}\rq\ along the HB are concerned,  
radiation-driven wind models for OB supergiants predict that 
the efficiency of mass loss jumps strongly by a factor of five 
at spectral type B1 (Vink et al. 1999, 2000). This is 
close to the position where the evidence for a gap in HB 
morphology is strongest. 
A mass-loss rate of the order of $10^{-10} - 10^{-11}$ \msunyr\ could be sufficient 
to explain the presence 
of the gap located at $\teff$\,$\simeq$\,20\,000 K; given an HB 
evolutionary timescale of $\approx10^8$ years with mass loss 
at this rate leads to a total 
amount of a few times $10^{-3}$ \msun\, sufficient to move an HB star 
by $\approx$1000 K, and so creating a \lq{gap}\rq. 

The above-mentioned issues, i.e. the presence of EHB stars, gaps, and 
anomalous abundances in HBs and sdB stars, prompted 
us to compute radiation-driven wind models for HB stars; to predict 
mass-loss rates for these objects, and subsequently explore 
their influence on evolutionary models. 
The mass loss computations may also provide valuable ingredients 
for HB angular momentum evolution and chemical separation calculations of 
sdB stars (see Unglaub \& Bues 2001). 

The outline of the paper is as follows. In the next section we describe 
the approach used for computing mass-loss rates, as well as the 
assumptions adopted in the numerical computations; in Sect.\,3 
we discuss the results concerning the mass-loss efficiency, where  
the dependence of \mdot\ on the main evolutionary parameters, the 
luminosity, effective temperature and stellar mass, as well as 
stellar metallicity, is presented. 
In Sect.\,4, we provide an analytical relation for \mdot\ 
as a function of the quoted parameters, which is useful for 
computing the mass-loss rates in evolutionary computations, and  
we investigate the effects of our recipe 
on HB stellar evolution (Sect.\,5). In Sect.\,6, we study the implications 
of mass loss regarding the ``zoo'' of problems in HB morphology that occur
for effective temperatures larger than $\simeq$\,10\,000 K, in particular the effects 
of mass loss on rotational velocities and the $\log g$ jump. 
Final remarks and conclusions will close the paper.

\section{Method and Assumptions in computing \mdot}

We compute mass-loss rates for HB stars 
under the hypothesis that radiation pressure on spectral lines
drives a stellar wind on the HB. 
Although this does not imply that there are no other physical 
mechanisms operating during the HB phase that could possibly drive 
a wind, it is the most sophisticated wind theory known, 
and it has been very successful in explaining the 
observed mass-loss rates of hot massive stars.

\subsection{The Monte Carlo method to calculate \mdot}
\label{s_method}

The description of the radiative wind driving 
with our method is based on a Monte Carlo technique 
that was first introduced by Abbott \& Lucy (1985).
This approach naturally accounts for photon-interactions 
with different metal ions, as the photons try 
to escape from the stellar wind.
In the Monte Carlo model used here ({\sc mc-wind}, de Koter et al. 1997, 
Vink et al. 1999), the momentum deposition is calculated using the 
Sobolev approximation by following the fate of a large number 
of photons that are released from below the stellar photosphere. 
To obtain a consistent solution, several wind models are calculated 
to find the mass-loss rate that is consistent with the radiative 
acceleration (see also Lucy \& Abbott 1993).

The calculation of radiation pressure with this method 
requires the input of a model atmosphere. The model atmospheres 
used in this study are the non-LTE unified Improved 
Sobolev Approximation code ({\sc isa-wind}), which treats  
the photosphere and wind in a unified manner (distinct from the 
so-called ``core-halo'' approaches).
For details of 
the code we refer the reader to de Koter et al. (1993,1997). 
The chemical species that are explicitly calculated in 
non-LTE are H, He, C, N, O, and Si. The iron-group elements 
are treated in a generalised version of the 
``modified nebular approximation'' (Lucy 1987, 1999).

\subsection{Assumptions in the models}

The model depends upon the assumption that 
the plasma behaves as a single fluid. As 
long as a large number of collisions between the
accelerating (C,N,O, and Fe-group) and 
non-accelerating (H and He) particles ensures  
a strong coupling, one can safely 
treat the wind as a single fluid.
A simple condition for this so-called 
\lq{Coulomb coupling}\rq~is given 
by Lamers \& Cassinelli (1999, p.\,193):

\begin{equation}
\frac{L_{*}~v}{\mdot} < 5.9~\times~10^{16}
\label{eq_cc}
\end{equation}
where $L_{*}$ is in $\lsun$, $v$ is in km\,${\rm s}^{-1}$ and 
$\mdot$ is in $\msunyr$.
Although this Coulomb condition is easily satisfied 
for the dense winds of massive O stars, it may break 
down for weaker winds, such as the winds from main-sequence A and B 
stars (Springmann \& Pauldrach 1992; Porter \& Drew 1995; 
Babel 1996; Krticka \& Kub{\'a}t 2000). However,
using typical values for an EHB star,  
log$L_{*}$\,=\,1.3; $v$\,=\,600 \kms; and log$\mdot$\,=\,$-11.70$ 
(see computations in Sect.\,\ref{s_massloss}), we find that the ratio 
($L_{*} v/\mdot$) from Eq.\,(\ref{eq_cc}) is ten times smaller than the quoted 
value of 5.9\,$\times$\,$10^{16}$, indicating that the condition of 
Coulomb Coupling is fulfilled.
It remains yet to be seen if the use 
of the Sobolev approximation is valid for weaker 
winds (see Owocki \& Puls 1999). 

\subsection{The adopted parameters}
\label{s_parameters}

\paragraph{Effective Temperatures~}
The models have effective temperatures between 
12\,500 and 35\,000 K with stepsizes of 2\,500 K.
We have checked this choice of gridding 
by computing additional models at intermediate temperatures,
which showed our initial resolution to be entirely 
adequate.  

\paragraph{Luminosities and Masses~}
The values for the stellar luminosity ($L$) and 
mass ($M$) were taken from evolutionary 
models (Cassisi \& Salaris 1997; Zoccali et al. 2000) 
for HB stars computed under different assumptions 
regarding the initial chemical composition.

\paragraph{Metal abundances~}
The adopted heavy elements distribution corresponds to the solar scaled
one provided by Anders \& Grevesse (1989), adopting $\zsun = 0.019$. 
For non-solar metallicity $Z$, the helium ($Y$) and hydrogen ($X$) 
abundances are adjusted in the following way: 

\begin{equation} 
Y~=~Y_{p}~+\left(\frac{\Delta Y}{\Delta Z}\right) Z
\label{eq_he}
\end{equation}
where the primordial helium abundance  
$Y_{\rm p}$\,=\,0.24 (Audouze 1987) and 
($\Delta Y/\Delta Z$)\,=\,3 (Pagel et al. 1992).
$X$ is then simply given by:

\begin{equation} 
X~=~1-~Y-~Z~
\end{equation}

\paragraph{Velocity Law~}
We calculated $\dot{M}$ for wind models with a 
$\beta$-type velocity law for the accelerating part 
of the wind:
\begin{equation}
\label{eq_betalaw}
v(r)~=~\vinf~\left(1~-~\frac{R_*}{r}\right)^\beta
\end{equation}
Below the sonic point, a smooth transition from this 
velocity structure is made to the velocity that follows 
from the photospheric density structure.
A value of $\beta=1$ was adopted in the accelerating 
part of the wind. This is a typical value for OB 
supergiants (Groenewegen \& Lamers 1989, Puls et al. 1996), but 
future mass loss observations of HB stars should 
be able to show whether or not this is also an adequate 
description for these low mass objects. Note that it has been  
demonstrated that over a $\beta$ range 0.7 -- 1.5 the $\dot{M}$ 
predictions are insensitive to the adopted value of $\beta$ 
(Vink et al. 2000).
We further assume a ratio \ratio\ =~1.0, as roughly  
indicated by $\vinf$ observations for a handful of SdO stars 
by Hamann et al. (1981) and Howarth (1987).


\section{The mass-loss predictions}
\label{s_massloss}

Using the procedure described in Sect.\,\ref{s_method}, we 
have calculated mass-loss rates as a function of $\teff$ with 
temperatures in the range between 12\,500 and 35\,000 K. 
This was performed for 
luminosities in the range log ($L_*/\Lsun$) 
= 1.3 -- 1.7 and masses in the range $M_*$ = 0.5 -- 0.7 
$\Msun$.

\subsection{Mass loss as a function of \teff}
\label{s_teff}

\begin{figure}
\centerline{\psfig{file=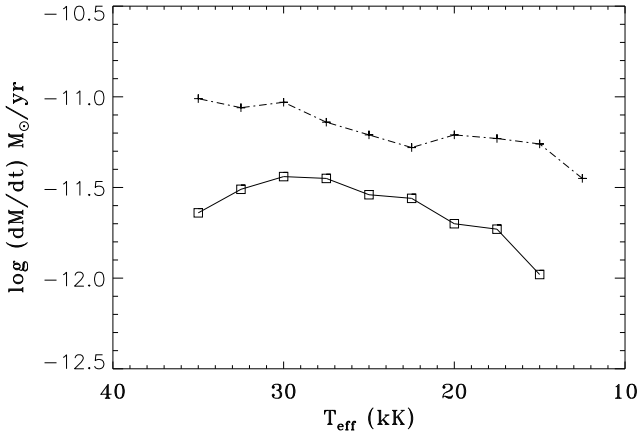, width = 9 cm}}
\caption{Mass loss predictions for HB stars as a function of effective temperature.
The solid line represents the computations for a solar metallicity. The dotted line is 
for a metallicity three times solar. The models are calculated for a constant mass 
of $M = 0.5 \msun$ and log$(L/\lsun) = 1.5$.}
\label{f_HBteff}
\end{figure}

The results of our predictions of HB mass loss as a function
of effective temperature are presented in Fig.\,\ref{f_HBteff}.
The solid line represents the computations for solar metallicity. To 
check the generality of this behaviour we have also computed mass loss
as a function of \teff\ for somewhat different input parameters, represented
by the dotted line, for a metallicity three times solar. In both cases there
appears to be a slight decrease of mass loss as a function of decreasing effective temperature. 
This can be attributed to the gradual shift of the flux maximum
towards longer wavelengths, and as the number of lines present
in the spectrum is smaller at higher wavelengths, the line
acceleration decreases, reducing the mass-loss
rate. Superimposed on this, one may have expected to see jumps,
where $\dot{M}$ could increase due to recombinations of
important line-driving ions. As mentioned earlier, for OB supergiants 
the mass loss increases steeply by a factor of five due to the recombination of 
Fe\,{\sc iv} to Fe\,{\sc iii} at spectral type B1.

Figure \ref{f_HBteff} however, indicates that these 
so-called ``bi-stability'' jumps are absent for HB stars. This is probably due to
the lower wind densities in HB stars in comparison to OB supergiants. 
In fact, these HB computations (for solar metallicity) are more comparable to the 
OB supergiant calculations at very low metallicities. Vink et al. (2001)
have shown that at lower wind densities, the winds are no longer driven
by Fe, but that the line driving by CNO-like elements takes over, and 
the dramatic recombinations are much less pronounced, or even absent 
at the temperatures under consideration here.

We conclude, that bi-stability jumps are not present for HB stars and we 
apply a fit through our computed datapoints over the complete temperature range. 
This temperature dependence of mass loss is later incorporated into 
our mass loss recipe for HB stars (Sect.\,\ref{s_recipe}).

\subsection{Mass loss as a function of $L$, $M$ and $Z$}
\label{s_other}

\begin{figure}
\centerline{\psfig{file=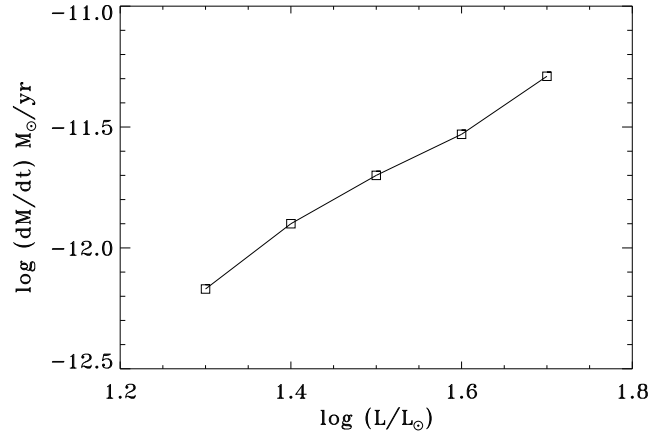, width = 9 cm}}
\caption{Mass loss predictions for HB stars as a function of stellar luminosity.
These models are calculated for a constant mass of $M = 0.5 \msun$, $\teff$\,=\,20\,000 K, and 
solar metallicity.}
\label{f_HBlogL}
\end{figure}

Predictions of HB mass loss as a function
of other stellar parameters, namely $L$, $M$, and $Z$ have 
also been performed. As an example, 
the results for mass loss as a function of $L$ are shown in 
Fig.\,\ref{f_HBlogL}.
%
%
%
%
%
The results of \mdot\ as a function of $L$, $M$ and $Z$ do not
yield any surprises. As expected, mass loss increases 
for increasing $L$ and $Z$, but decreasing $M$. Although the 
behaviour is qualitatively similar to the O star recipe in Vink et al. (2000), 
the dependencies are somewhat different. The values of 
these dependencies and  
mass-loss rates are discussed in the next section.


\section{Mass loss recipe}
\label{s_recipe}

In this section, we present a mass loss recipe for HB stars as a function
of basic stellar parameters. 
To obtain the recipe we have determined four separate dependencies, and  
checked if they were independently applicable. As this was found to be the 
case, we have combined the four independent parameters 
from Sects.\,\ref{s_teff} and \ref{s_other} and constructed the 
following analytical relationship for HB mass loss: 

\begin{eqnarray}
{\rm log}~\dot{M} & = &~-11.70~(\pm 0.08) \nonumber \\
                  & &~+~1.07~(\pm 0.32)~{\rm log} (\teff/20000) \nonumber\\
                  & &~+~2.13~(\pm 0.09)~({\rm log}L_* - 1.5) \nonumber\\
                  & &~-~1.09~(\pm 0.05)~{\rm log}(M_*/0.5) \nonumber\\
                  & &~+~0.97~(\pm 0.04)~{\rm log}(Z_*) \nonumber\\
                  \nonumber\\
                  & &~{\rm derived~for:} \nonumber\\
                  & &              \,12\,500 \le \teff \le 35\,000\,{\rm K} \nonumber\\
                  & &                  \,1.3 \le {\rm log}L_*  \le 1.7 \nonumber\\
                  & &                   \,0.5 \le {M_*} \le 0.7 \nonumber\\
                  & &                   \,0.1 \le {Z_*} \le 10 \nonumber\\
\label{eq_HBfit}
\end{eqnarray}
where \Teff\ is in Kelvin and $L_*$, $Z_*$, and $M_{*}$ are all given 
in solar units. 
Note that the computations have been performed for the case 
$\vinf\,=\,\vesc$ (see Sect.\,\ref{s_parameters}). As discussed in 
Vink \& de Koter (2002), \mdot\ is maximal for this value, and  
Eq.\,(\ref{eq_HBfit}) therefore provides an upper limit to the 
mass-loss rate. 
In case \vinf\ is not equal to \vesc, one can use the (\ratio) dependence 
of \mdot\ as derived in Vink et al. (2000) for OB stars.
The root-mean-square (rms) difference between the results following 
from Eq.\,(\ref{eq_HBfit}) and the actual model computations is 0.08 dex 
in log $\dot{M}$.
This implies that the HB formula yields a good representation of 
the actual HB mass loss computations. 

Alternatively, if we put the HB stellar parameters into the recipe 
of Vink et al. (2000) for massive O stars (and keep \vinf\ fixed to \vesc), 
we find an rms difference of 0.44 dex in the mass loss, leading to the 
conclusions that an extrapolation of the O star recipe by four orders of magnitude in 
stellar luminosity would have resulted in mass-loss rates that 
are systematically too high by about a factor of two. 
One may still wonder whether the HB mass loss recipe of Eq.\,(\ref{eq_HBfit}) is also applicable 
to stars with stellar parameters for which the recipe was not specifically derived.
To check whether Eq.\,(\ref{eq_HBfit}) may safely be used over a wider range in stellar parameters, we have 
performed mass loss calculations for the winds of post-HB stars and compared 
these actual calculations with the results from the recipe. We find that 
the mass loss recipe may be used for other classes of low-mass blue objects 
as well, as long as the desired accuracy is within a factor of $\sim$\,2. 
As there are hardly any mass-loss predictions available for these types 
of objects either, the HB formula may be applied to all hot, low-mass stars,  
of the types: HB, EHB, sdB, sdOB, post-HB, AGB-manqu{\'e}, UV-bright stars, 
and extreme helium stars, as long as their effective temperatures 
are not significantly higher than $\sim$\,40\,000 K. This because existing 
mass loss calculations, such as the ones presented here, but also those 
by Pauldrach et al. (1988) for Central Stars of Planetary Nebulae have the 
problem that line lists become incomplete with respect to higher ionisation stages. 
Note that we do not expect problems with extrapolating Eq.\,(\ref{eq_HBfit}) 
to effective temperatures as low as $\approx\,8\,000$\,K.   
A computer routine of the HB mass loss recipe is available upon
request or on the Web.\footnote{http://astro.ic.ac.uk/$\sim$jvink/} 

\section{Mass loss effects on HB evolution}
\label{s_evolution}

\subsection{Assumptions in Evolutionary Tracks and Outer Boundary conditions}

In order to check the effects of the computed mass-loss recipe 
on evolutionary tracks of HB stars, we have computed 
two series of models, identical in every way, except that one set incorporates the 
mass-loss recipe, while the occurrence of mass loss is 
neglected in the other.
All models have been computed using the {\sc franec} evolutionary code
(Cassisi \& Salaris 1997, Castellani et al. 1997 and references therein). 
As far as the adopted physical 
input parameters as well as the treatment of convection 
during the central He-burning phase are concerned, 
we refer the interested reader to the papers by 
Cassisi \& Salaris (1997) and Zoccali et al. (2000). 
The treatment of outer boundary conditions is performed, as usual, by 
adopting a $T(\tau)$ relation (Krishna-Swamy 1966). 

\begin{figure*}
\sidecaption
\includegraphics[width=12.0cm]{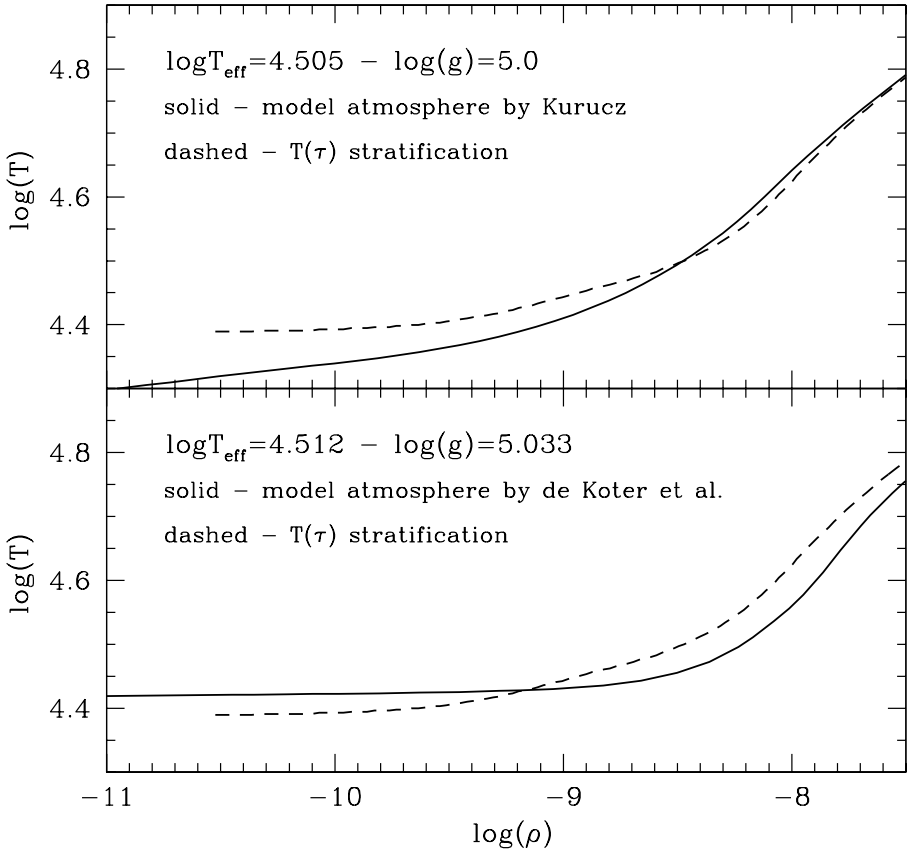}
\caption{{\sl Upper panel:} Comparison between the atmosphere thermal stratification provided 
by the $T(\tau)$ integration and by the model atmosphere of Kurucz (1993) for the labelled
values of effective temperature and gravity. {\sl Bottom panel:} As upper panel, but for the
model atmospheres of de Koter et al. (1993, 1997), for the labelled values of 
effective temperature and gravity. In all cases, a chemical composition of 
$Z$\,=\,0.012\,; $Y$\,=\,0.25 has been adopted.}
\label{atmstrut}
\end{figure*}

To check the validity of this assumption for hot HB stars, we have computed 
a large grid of model atmospheres that provide more accurate 
descriptions of the thermal stratification of the atmospheres of these 
stars, and investigated whether a different treatment of the outer boundary 
conditions influences the evolutionary output parameters. 
To this end, we computed several HB models for different 
metallicities by adopting in one case the $T(\tau)$ relation from Krishna-Swamy, and in 
the other the boundary conditions provided by the more sophisticated model
atmosphere computations. The model atmospheres used for this test 
were the non-LTE {\sc isa-wind} models of de Koter et al. (1993, 1997) with 
negligible mass loss, as well as the hydrostatic LTE, line-blanketed models 
of Kurucz (1993). In the model atmosphere cases, the connection between the atmospheres and 
the internal structures has been fixed at $\tau_{\rm Ros} = 10$. 
Note that we have verified that the results obtained are not affected by this choice 
of matching point between stellar atmosphere and internal regions.
Our numerical experiments clearly showed that the stellar
properties such as the effective temperature, are not significantly 
affected by the assumptions made concerning the outer boundary conditions. This is
because the $T(\tau)$ approach provides an estimate of the thermal
stratification in the stellar atmosphere that is in good agreement with 
those provided by more accurate model atmospheres computations, as can be seen 
in Fig.\,\ref{atmstrut}, where we have plotted the thermal stratifications 
provided by the $T(\tau)$ method, and the model atmospheres quoted above, 
for values of the effective temperature and surface gravity suitable 
for hot HB stars. These results imply that canonical theoretical predictions 
of effective temperatures of hot HB stars can be considered to be robust.

\subsection{HB Evolution with Mass Loss for metal rich clusters}

In order to maximise the effects of mass loss on the HB evolution, and given that 
mass loss increases with stellar metallicity, we have computed HB models 
for a metallicity twice solar\footnote{All evolutionary
models computed here as well as additional models can be obtained upon 
request}. 
This choice allows a direct comparison with the evolutionary 
computations performed by Yong et al. (2000). 
All HB models have a 1 \Msun\ RGB progenitor with an initial chemical
composition of $Z$\,=\,0.04 and $Y$\,=\,0.34. 
The He core mass of this structure at He ignition is equal 
to 0.466 \Msun, while its surface He abundance in the same 
evolutionary phase is of the order of $Y$\,=\,0.36. 
The standard models, i.e. the ones computed neglecting mass loss,
have been presented by Bono et al. (1997).

\begin{figure*}
\centerline{\psfig{file=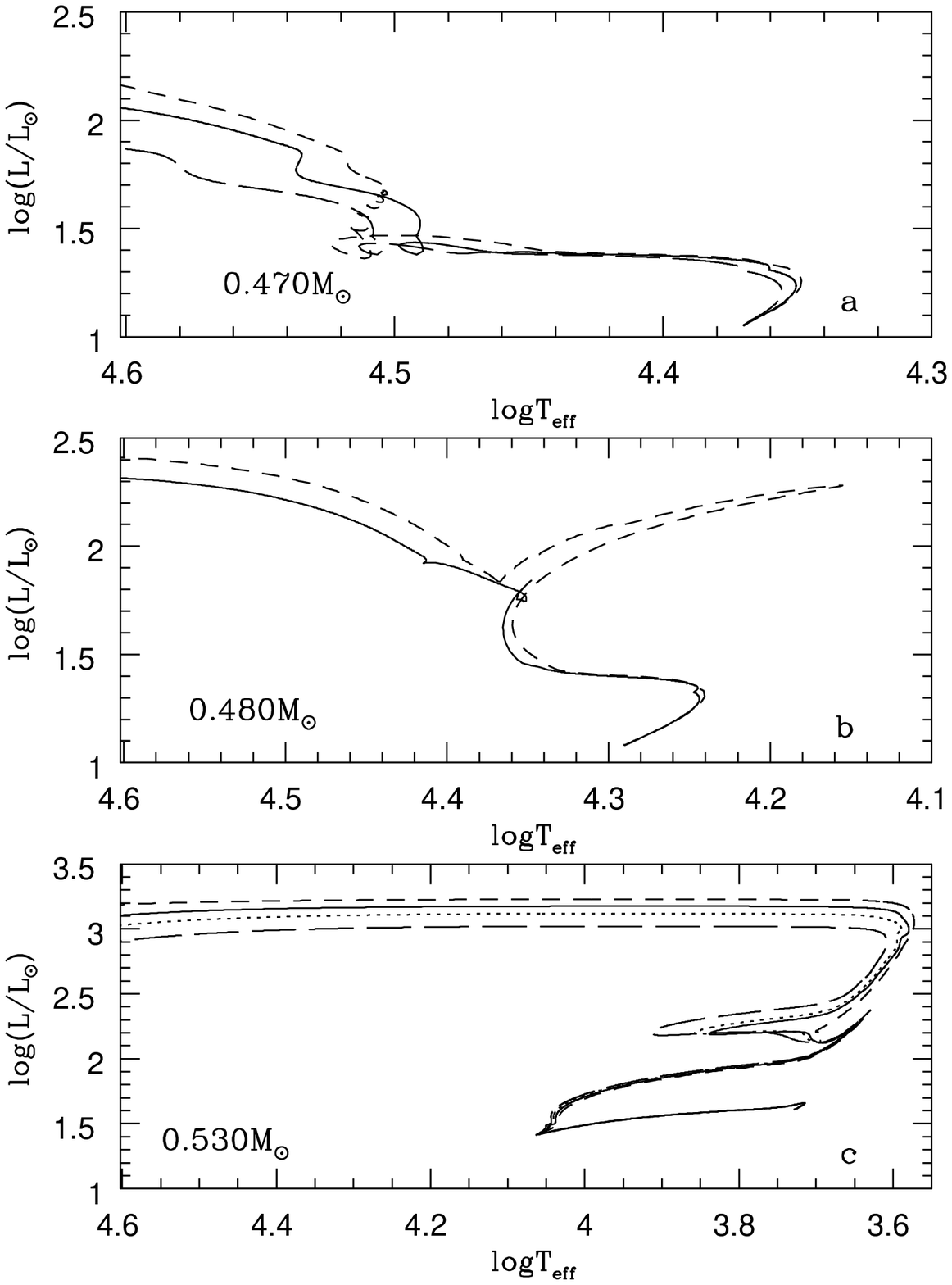, width = 16 cm, clip=}}
\caption{The H-R diagram representing the evolution of selected HB models 
with various initial mass (as labelled) whose RGB progenitor mass 
is equal to 1 $M_\odot$; with an initial chemical composition of $Z\,=\,0.04$\,; $Y\,=\,0.34$. 
In each panel, the short dashed line corresponds to the standard model (neglecting mass loss), while
the solid line refers to models accounting for mass loss according to Eq.\,(\ref{eq_HBfit}). 
In panels (a) and (c), models accounting for an enhancement of the surface metallicity
(see text for details) are also plotted (dotted line -- $Z\,=\,0.1$; long-dashed line -- $Z\,=\,0.2$).}
\label{evolo}
\end{figure*}

In the various panels of Fig.\,\ref{evolo}, a comparison 
between standard models and models accounting for mass loss 
(according to Eq.\,\ref{eq_HBfit}) is presented. 
When computing the models with mass loss, we have accounted for 
this process along the whole evolutionary path, starting  
from the Zero Age Horizontal Branch (ZAHB) until an effective temperature of the order of
40\,000K (see previous discussion). This implies that we are also using 
our mass loss recipe out of its validity range in luminosity. 
In the case of the most massive star, i.e. the coolest one, with a  
ZAHB location below 10\,000 K, we are also slightly extrapolating Eq.\,(\ref{eq_HBfit}) 
out of its validity range in \teff.

We would expect that the computed mass-loss rates of the order 
of $10^{-12}$ \msunyr are too low to alter 
evolutionary tracks in a major way. Indeed, as shown in Fig.\,\ref{evolo}, the evolutionary 
paths of the selected models are not significantly affected by the occurrence 
of mass loss at the computed rates. 
The less massive, hottest model has lost an amount of mass of 
$2.2\cdot10^{-4}M_\odot$ at the end of the 
He central burning phase, while the coolest model, the one 
with mass equal to $0.530M_\odot$, has lost 
$2.4\cdot10^{-4}M_\odot$ at the end 
of the same evolutionary phase.
The amount of mass lost is slightly 
larger for the more massive model since this model is brighter, 
and the mass-loss rate most strongly depends on $L$. 
From these evolutionary computations, we arrive at the following conclusions:

\begin{itemize}

\item{} Mass-loss rates due to radiation pressure are not sufficient as 
to significantly modify evolutionary tracks for HB stars.

\item{} The mass-loss computations do not provide the high rates needed 
in the Yong et al. (2000) scenario for explaining the occurrence of
EHB stars in metal-rich clusters such as NGC\,6791.

\item{} Since the efficiency of radiation-driven wind
decreases with stellar metallicity, the effects are likely to be even
smaller for HB stars in more metal-poor clusters. 


\end{itemize}

\subsection{HB Evolution with increasing Mass Loss due to radiative levitation}

Before we present the effects of increased mass loss due to 
radiative levitation on HB evolution, we first discuss the
connection between mass loss and the changes of the 
surface chemical abundances due to these physical processes of 
radiative levitation and atomic diffusion. 
Our calculations show that the mass-loss efficiency increases strongly 
when the envelope abundances of heavy elements increase as a consequence 
of radiative levitation. Nevertheless, it is also well
known that mass loss works as a competing process to diffusion by decreasing 
the efficiency of radiative levitation in producing large chemical 
overabundances -- at least for heavy elements such as silicon. 
In particular, Michaud \& Charland (1986) have shown that if mass loss 
increases beyond $10^{-14}$ \msunyr, chemical overabundances could be wiped out.
In more recent times, Unglaub \& Bues (2001) have investigated the influence of
diffusion and mass loss on the chemical composition of sdB stars. The main
outcome of their work was that observed chemical patterns can only be explained 
if mass-loss rates are in the range $10^{-14}\,<\,$\Mdot(\msunyr)$\,<\,10^{-12}$. 
Higher rates would basically prevent the effects of diffusion, whereas for 
lower rates helium would sink in too short time scales 
compared to the typical lifetime of an sdB star. Note that 
our mass-loss predictions fall in the middle of the range of the calculations by   
Unglaub \& Bues (2001) for metallicities typical of GCs with extended 
blue tails; once radiative levitation becomes effective in strongly 
increasing the stellar metallicity, mass loss strongly increases, and becomes 
of the same order of magnitude to -- or even larger than -- 
the upper limit quoted in the work of Unglaub \& Bues. 
If these physical processes indeed occur in real stars, they 
should have strong effects on the measured abundances of helium and 
heavy elements in hot HB stars. A detailed investigation into this 
topic is beyond the scope of the present work, but we wish to emphasise the 
further need for accurate chemical separation computations that consistently 
account for diffusion, radiative levitation, and reliable mass-loss predictions. This, 
as a function of metallicity and other stellar parameters such as mass, effective 
temperature and luminosity.

We now turn to the effect of increased mass loss on the evolution
of HB stars. Given the importance of the metallicity on the line driving 
efficiency, we maximise the effects of radiative levitation
on the surface abundance. This was done by simply assuming 
that starting from the ZAHB the surface metallicity is               
equal to five times or ten times the solar value. 
Note that we do not properly account for radiative levitation in the 
evolutionary code, but assume that the diffusion process is a fast 
and efficient mechanism to increase the surface metallicity (Michaud et al. 1985).
Although this is a rather crude approach, we are only interested in checking 
the maximum effect of radiative levitation on the mass-loss efficiency, and 
we do therefore not account for possible changes in the envelope opacity 
properties due to this metallicity increase. In case we would have accounted 
for these effects, the models would have been somewhat fainter and cooler, which 
would have had the effect of slightly decreasing the mass-loss efficiency. 
In other words, our more crude approach has the effect of maximising 
the efficiency of mass loss. 

The additional experiments (also shown in Fig.\,\ref{evolo}), clearly show 
that even with the assumed increase of the surface stellar metallicity, 
the mass-loss efficiency is still too low to affect the HB evolution.
Therefore, it is worth noting that:
\begin{itemize}

\item{} It is almost impossible that an {\it in situ} mass loss 
process due to radiation pressure alone can affect 
the HB mass distribution in any major way.

\item{} A relevant parameter in population synthesis studies is the maximum 
mass of HB structures that at the end of the central He-burning phase behave as 
AGB-manqu{\'e} stars (Greggio \& Renzini 1990). Due to the low efficiency of mass 
loss during the HB phase, this critical mass is not affected by the inclusion of 
our mass loss recipe in stellar model computations.

\end{itemize}

\section{Mass loss, HB rotation rates, \& the $\log g$ jump}
\label{s_hbmor}

In this section we discuss the implication of mass loss 
on HB rotational velocities and the ``low gravity'' problem 
for hot HB stars, which is represented by a jump in $\log g$ at 
$\teff$\,$\simeq$\,11\,000 K.

Observational analyses of rotational velocities of HB stars 
show that among the cool group ($\teff$\,$\le$\,11\,000 K) 
both fast and slow rotators are present (Peterson et al. 1995), but that 
for the hot group ($\teff$\,$>$\,11\,000 K) {\it all} stars 
rotate slowly (Behr et al. 2000a, 2000b ; Recio-Blanco et al. 2002). 
We argue that the absence of fast rotators
in HB stars hotter than 11\,000 K can be explained by a stellar wind
set up by radiative levitation of heavy elements (Sweigart 2000), which could
contribute significantly to the removal of angular momentum. 
Recio-Blanco et al. (2002) argue that the mass-loss rate may increase 
by a large factor between 10\,000 and 20\,000 K due to a change in the 
ionisation state of hot star winds (referring to the work by 
Vink et al. 2000 on massive stars). However, as we have 
shown in Fig.\,\ref{f_HBteff}, dramatic changes in the mass-loss rate 
over $\teff$ are absent for HB stars.  
This notwithstanding, as can be noted from Eq.\,(\ref{eq_HBfit}), mass loss 
increases by about 2 dex when the photospheric metal abundance ($Z$) 
increases by 2 dex. We therefore argue that the striking photospheric 
abundances in hot HB stars, which are most likely due to the onset of 
radiative levitation, may spin down the surface velocities of HB stars hotter 
than 11\,000 K, explaining the absence of fast rotation at these temperatures. 
Note that firmer and more 
quantitative conclusions can only be achieved by understanding the coupling 
between mass loss and stellar rotation (see Soker \& Harpaz 2000).

In addition, we question whether the increase in the mass-loss rate around 
11\,000 K invalidates the use of hydrostatic model atmospheres (such as 
those by Kurucz) for hot HB stars. 
For massive O stars it is a well-known fact that neglecting winds 
in model atmosphere calculations causes systematic errors in the derived 
atmospheric parameters, notably $\log g$. 
This has even led to a systematic discrepancy between masses derived from 
stellar spectra vs. those from evolutionary models, 
the ``mass-discrepancy'' (Groenewegen \& Lamers 1989; 
Herrero et al. 1992). Indeed, the luminosities for HB stars are much 
lower (having the effect of lowering the mass-loss rates), but the stars 
are also less massive (increasing $\dot{M}$), and have much smaller radii, 
which substantially increases the more relevant parameter, the mass flux. 
It is therefore not at all obvious that hydrostatic model atmospheres are 
applicable to these types of objects. 
Even more so, atmospheric analyses for hot HB stars have shown that there is also 
a ``mass discrepancy'' for these objects (Moehler et al. 1995), whereas 
the atmospheric determinations and canonical evolutionary models do 
agree for the cooler HB stars. 
Although the ``low gravity'' problem for hot HB stars may partially 
be explained by radiative levitation of metals (Grundahl et al. 1999; 
Moehler et al. 2000), it has not been solved completely (Moehler 2001), 
the residual discrepancy that is still left between $15\,000\,<\,\teff\,<\,20\,000\,$K
is about 0.1 dex in $\log{g}$. 

To test whether stellar winds have a noticeable effect on stellar spectra of 
hot HB stars, we compute H$\gamma$ line profiles for these objects 
using the model atmosphere code {\sc isa-wind} and the synthetic 
spectrum code {\sc wynspec} (de Koter et al. 1997).  
We compare H$\gamma$ line profiles for a solar metallicity star with 
the following stellar parameters: $\teff$\,=\,17\,500 K, 
$M$\,=\,0.56\,$\msun$, and $\log(L/L_\odot)$\,=\,1.37, corresponding 
to $\log g$\,=\,4.73. Using Eq.\,(\ref{eq_HBfit}), we expect such a star to have a mass-loss rate 
$\log \mdot$ ($\msunyr$)\,=\,$-$11.85, and we therefore use this value in the computation of 
an H$\gamma$ line profile, and compare this line profile to a model with negligible 
mass loss (we choose a rate typical for the Sun: $\log\mdot$\,($\msunyr$)\,=\,$-$14).

\begin{figure}
\centerline{\psfig{file=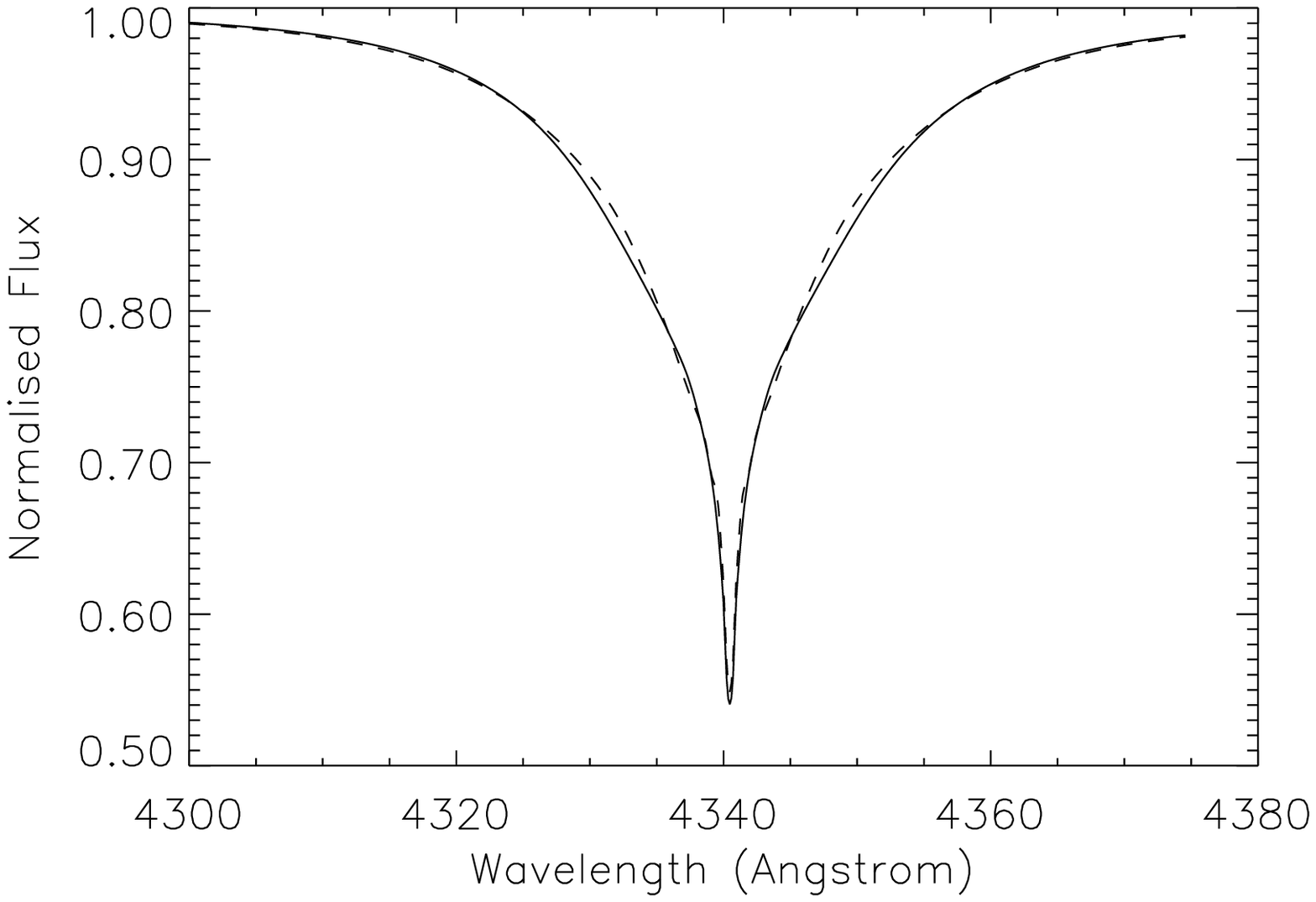, width = 9 cm}}
\centerline{\psfig{file=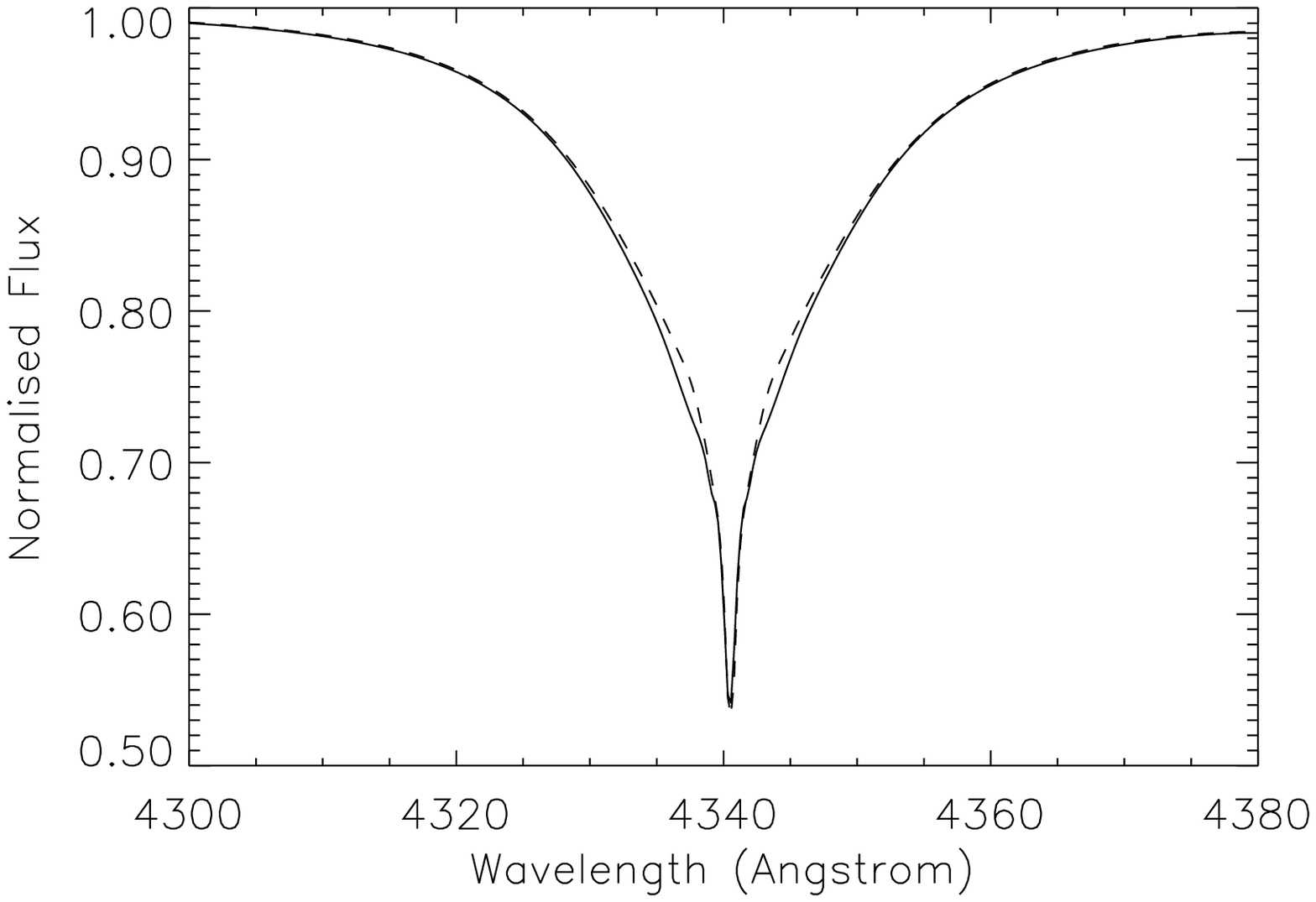, width = 9 cm}}
\caption{{\sl Upper panel:} the normalised flux at H$\gamma$ for a model with 
an expected mass-loss rate of $\log\mdot(\msunyr)$\,$= -$11.85 (dashed line), and 
a model for negligible mass loss, i.e. $\log\mdot(\msunyr)$$\simeq -$14 (solid line).
{\sl Bottom panel:} a comparison for the high gravity ($\log g$\,=\,4.73; solid line) and 
the low gravity ($\log g$\,=\,4.63; dashed line) model.
The other stellar parameters in both panels are: $\teff$\,=\,17\,500 K, 
$\log(L/L_\odot)$\,=\,1.37, and solar metallicities.}
\label{f_comp}
\end{figure}

As can be seen in the upper panel of Fig.\,\ref{f_comp} wind emission in the 
model with significant mass loss (dashed line) has a noticeable effect on the 
line wings, which may mimic a lower $\log g$ from Balmer  H$\gamma$ line 
profiles. To check whether the $\log g$ jump could indeed be an artifact 
of the use of hydrostatic model atmospheres, we compare H$\gamma$ line profiles 
for models with different $\log g$ values: $\log g$\,=\,4.73, vs. a model with $\log g$\,=\,4.63 (or 
equivalently a mass of $M$\,=\,0.44\,$\msun$), and keep all other parameters fixed.
This comparison is also shown in Fig.\,\ref{f_comp}.
The bottom panel shows H$\gamma$ for the high gravity (solid line) and 
the low gravity (dashed line) model. As expected, Stark effects broaden the line 
wings for the higher gravity model.

Although the effects in the two panels of Fig.\,\ref{f_comp} are small, they yield the same 
systematic trend. They are so similar that it suggests that the neglect of winds in 
atmospheric analyses of hot HB stars can mimic too low surface gravities. An increase of 
mass loss due to the increased surface abundances of hot HB stars may therefore  
invalidate the use of hydrostatic atmospheres for these stars. Note that the use of 
such models is considered robust for the cooler HB models, where indeed no discrepancy between
evolutionary and spectroscopic masses has been reported. 
We argue that the ``wind emission'' effect -- together with the 
metal enhancement due to radiative levitation -- is the most 
likely explanation for the observed jump in $\log g$.

We then arrive at the following scenario for the ``zoo'' of problems 
in HB morphology around 11\,000K:
due to the more stable atmospheres of the hotter HB stars 
radiative levitation increases the metal abundances for these objects.
This subsequently explains: (i) the striking surface abundances in 
hot HB stars; (ii) the existence of the Str{\"o}mgren u-jump. It also 
reduces the $\log g$ discrepancy.
As the abundance spectroscopic determinations indicate that only the hotter HB 
models are affected by radiative levitation, a stellar wind is expected 
to be set up with an increased $\dot{M}$ by two 
orders of magnitude. This onset of (significant) mass loss can then
also help in explaining (iii) the absence of fast rotators for HB stars, 
and (iv) the jump in $\log g$. In passing, we finally note that (v) the gap 
observed along the HB of many GCs at $(B-V)\simeq0$ ($\teff$\,$\simeq$\,9\,000 K) could also be 
due to an atmospheric phenomenon related to the chemical peculiarities induced by 
radiative levitation as suggested by Caloi (1999).

\section{Conclusions \& Outlook}
\label{theend}

In this paper we have, for the first time, computed mass-loss rates for 
HB stars. We have shown that the computed rates, as predicted by the most plausible 
mechanism of radiation pressure on spectral lines, are too low to produce EHB/sdB 
stars. This invalidates the scenario outlined by Yong et al. (2000) 
to create these objects by excessive mass loss {\it on} the HB. 
We argue, however, that mass loss plays a role in the distribution of 
rotational velocities of hot HB stars, and for the so-called
$\log g$ jump. The mass loss recipe derived in this paper is, strictly speaking, 
only valid for HB stars, but as there are hardly any mass-loss predictions 
available for low-mass blue stars, the recipe may also be applied to: post-HB, 
AGB-manqu{\'e}, UV-bright stars, extreme helium stars, as long as 
the desired accuracy is within a factor of two, and as long as the 
effective temperatures are not higher than 40\,000 K.

Although we have proposed a scenario where winds are ubiquitous for hot HB stars, 
and subsequently affect the rotational 
velocities, as well as the atmospheric parameters ($\log g$), there is still a lot of work 
to be done.

First and foremost, spectral evidence for mass loss in HB and sdB stars ought to 
be sought to check whether the mass-loss rates, as derived in this paper, indeed occur. 
Diffusion calculations including mass loss for sdB stars (Unglaub \& Bues 2001) suggest 
that our derived mass-loss rates are reasonable, but this is certainly 
not a model-independent check. 

Second, evolutionary models including rotation (first steps have been undertaken
by Sills \& Pinsonneault 2000) and mass loss should be 
computed to see whether the 
absence of fast rotators for stars hotter than 11\,000 K can indeed be due to the 
removal of angular momentum due to stellar winds.

Last but not least, systematic atmospheric analyses of hot HB stars accounting for the 
{\sl actual} surface heavy elements distribution and including 
mass loss should be performed to see whether the $\log g$ jump is indeed an artifact
of the adopted hydrostatic model atmospheres. The current situation, where evolutionary 
models are not in agreement with the spectral analyses is highly undesirable, as this 
suggests that current stellar evolution theory is not only incapable of producing 
extreme HB stars, but that even ``normal'' blue HB stars pose a serious problem.
In other words, a solution to the ``low gravity'' problem for hot HB stars 
could significantly enhance our current understanding of the later phases of 
stellar evolution.


\begin{acknowledgements}

We would like to thank Allen Sweigert and Marcio Catelan for constructive 
comments that helped improve the paper. We would additionally like to 
thank Giuseppe Bono and Rubina Kotak for fruitful discussions, Noam Soker 
for a swiftly produced referee report, and Alex de Koter and Robert Kurucz 
for the use of their model atmospheres. 
SC acknowledges financial support by MURST-Cofin2002.

\end{acknowledgements}


\begin{thebibliography}{}

\bb
Abbott, D.C., \& Lucy, L.B., 1985, ApJ 288, 679

\bibitem[Anders \& Grevesse 1989]{anders89}
Anders, E., \& Grevesse, N., 1989, Geochim. Cosmochim. Acta 53, 197

\bibitem[Audouze 1987]{audou87}
Audouze, J., 1987, Observational Cosmology, IAU Symp. 124, eds. A. Hewitt et al., 
Reidel Publ. p. 89

\bb
Babel, J., 1996, A\&A 309, 867

\bb
Behr, B.B., Cohen, J.G., McCarthy, J.K., \& Djorgovski, S.G., 1999, ApJ 517, L135

\bb
Behr, B.B., Djorgovski, S.G., Cohen, J.G., et al. 2000a, ApJ 528, 849

\bb
Behr, B.B., Cohen, J.G., \& McCarthy, J.K., 2000b, ApJ 531, L37

\bb
Bergeron, P., Wesemael, F., Michaud, G., \& Fontaine, G., 1988, ApJ 332, 964

\bb
Bono, G., Caputo, F., Cassisi, S., Castellani, V., \& Marconi, M., 1997, ApJ 489, 822

\bb
Brown, T.M., Sweigart, A.V., Lanz, T., Landsman, W.B., \& Hubeny, I., 2001, ApJ
562, 368

\bb
Caloi, V., 1999, A\&A 343, 904

\bb
Cassisi, S., \& Salaris, M., 1997, MNRAS 285, 593

\bb 
Castellani, V., Ciacio, F., Degl'Innocenti, S., \& Fiorentini, S., 1997, AAP 322, 801

\bb
Castor, J.I., Abbott, D.C., \& Klein, R.I., 1975, ApJ 195, 157

\bb
Catelan, M., Borissova, J., Sweigart, A.V., \& Spassova, N., 1998, ApJ 494, 265

\bb
Catelan, M., Bellazzini, M., Landsman, W.B., et al. 2001, AJ 122, 317

\bb
de Koter, A.,Schmutz, W., \& Lamers, H.J.G.L.M., 1993, A\&A 277, 561

\bb
de Koter, A., Heap, S.R., \& Hubeny, I., 1997, ApJ 477, 792

\bb
Dorman, B, O'Connell, R.W., \& Rood, R.T., 1995, ApJ 442, 105

\bb
Ferraro, F.R., Paltrinieri, B., Fusi Pecci, F., Rood, R.T., \&  Dorman, B., 1998, ApJ, 500, 311

\bb
Fusi Pecci, F., Ferraro, F.R., Bellazzini, et al. 1993, AJ 105, 1145

\bb
Glaspey, J.W., Michaud, G., Moffat, A.F.J., \& Demers, S., 1989, ApJ 339, 926

\bb
Green R.F., Schmidt M., \& Liebert J., 1986, ApJS 61, 305

\bb
Greenstein, J.L., 1971, Proceedings from IAU Symposium n. 42., W.J. Luyten eds., 
Springer-Verlag, Dordrecht, p.\,46

\bb
Greggio, L., \& Renzini, A., 1990, ApJ 364, 35

\bb
Groenewegen, M. A. T., \& Lamers, H. J. G. L. M., 1989, A\&AS 79, 359

\bb
Grundahl, F., Catelan, M., Landsman, W.B., Stetson, P.B., \& Andersen, M.I., 1999,
ApJ 524, 242

\bb 
Hamann, W.-R., Gruschinske, J., Kudritzki, R. P., \& Simon, K. P., 1981, A\&A 104, 249

\bb 
Heber, U., Moehler, S., Napiwotzki, R., Thejll, P., \& Green, E.M., 2002, A\&A,
383, 938

\bb
Herrero A., Kudritzki, R. P., Vilchez, J. M, et al. 1992, A\&A 261, 209

\bb
Howarth, I.D., 1987, MNRAS 225, 33

\bibitem[]{}
Kilkenny, D., Koen, C., O'Donoghue, D., \& Stobie, R.S., 1997, MNRAS 285, 640

\bb 
Koopmann, R.A., Lee, Y,-W., Demarque, P., \& Howard, J.M., 1994, ApJ 423, 380

\bb
Krishna-Swamy, K. S., 1966, ApJ, 145, 174

\bb
Krticka, J., \& Kub{\'a}t, J., 2000, A\&A 359, 983

\bb
Kurucz, R.L. 1993, SAO CD-ROM   

\bb
Lamers, H.J.G.L.M., \& Cassinelli, J.P., 1999, Introduction to stellar winds, Cambridge Univ. Press

\bb
Lee, Y-.W., Demarque, P., \& Zinn, R., 1994, ApJ 423, L248

\bibitem[]{}
        Lucy, L.B., 1987, In: ESO Workshop on the SN 1987A, 
        Proceedings (A88-35301 14-90), 417

\bibitem[]{}
        Lucy, L.B., 1999, A\&A 345, 211

\bibitem[Lucy \& Abbott 1993]{lucy93}
Lucy, L.B., \& Abbott, D.C., 1993, ApJ 405, 738

\bb
Lucy, L.B., \& Solomon, P., 1970, ApJ 159, 879

\bb
Mengel, J.G., Norris, J., \& Gross, P.G., 1976, ApJ 204, 488

\bb
Michaud, G., \& Charland, Y., 1986, ApJ, 311, 326

\bb
Michaud, G., Bergeron, P., Wesemael, F., \& Fontaine G., 1985, ApJ 299, 741

\bb
Moehler, S., 2001, PASP 113, 1162

\bb
Moehler, S., Heber, U., \& de Boer, K.S., 1995, A\&A 294, 65

\bb
Moehler, S., Sweigart, A.V., Landsman, W.B., \& Heber, U., 2000, A\&A 360, 120

\bb
Newell, E.B., 1973, ApJS 26, 37

\bibitem[Owocki \& Puls 1999]{owocki99}
        Owocki S.P., \& Puls J., 1999, ApJ 510, 355

\bibitem[Pagel et al. 1992]{pagel92}
Pagel, B.E.J., Simonson, E.A., Terlevich, R.J., \& Edmunds, M.G., 1992, MNRAS 255, 325 

\bibitem[Pauldrach et al. 1988]{paul88}
Pauldrach, A.W.A., Puls, J., Kudritzki R.P., M{\'e}ndez, R.H., \& Heap, S.R., 1988, A\&A 207, 123

\bb
Peterson, R.C., Rood, R.T., \& Crocker, D.A., 1995, ApJ 452, 214

\bb
Piotto, G., Zoccali, M., King, I.R., et al. 1999, AJ 118, 1727

\bb 
Porter, J.M., \& Drew, J.E., 1995, A\&A 296, 761

\bb
Puls J., Kudritzki R.-P., \& Herrero A., 1996, A\&A 305, 171

\bb
Raimondo, G., Castellani, V., Cassisi, S., Brocato, E., \& Piotto, G., 2002,
ApJ 569, 975

\bb
Recio-Blanco, A., Piotto, G., Aparicio, A., \& Renzini, A., 2002, ApJL {\sl in
press}, (astro-ph/0204403)

\bb
Rich, R.M., Sosin C., Djorgovski, S., et al. 1997, ApJ 484, L25

\bb
Sandage, A., \& Wallerstein, G., 1960, ApJ 131, 598

\bb
Sandage, A., \& Wildey, R., 1967, ApJ 150, 469

\bb
Sills, A., \& Pinsonneault, M. H., 2000, ApJ 540, 489

\bb
Soker, N., 1998, AJ 116, 1308

\bb
Soker, N., \& Harpaz A., 2000, MNRAS 317, 861

\bb 
Soker, N., Catelan, M., \& Rood, R.T., Harpaz A., 2001, ApJ 563, L69

\bb 
Springmann, U. W. E., \& Pauldrach, A. W. A. 1992, A\&A 262, 515


\bb 
Sweigart, A.V., 1997, ApJ 474, L23

\bb 
Sweigart, A.V., 2000, in: ``Mixing and Diffusion in Stars: Theoretical Predictions 
                and Observational Constraints'', 24th meeting of the IAU, Joint
		Discussion , (astro-ph/0103133), {\sl in press}
		
\bb
Sweigart, A.V., Brown, T.M., Lanz, T., Landsman, W.B., \& Hubeny, I., 2002,
Proceeding of the Workshop "Omega Centauri: a Unique Window into Astrophysics" 
(Cambridge, August, 2001), ASP Conf. Ser., edited by F. van Leeuwen, G. Piotto, and J. 
Hughes, {\sl in press}

\bb
Unglaub, K. \& Bues, I., 2001, A\&A 374, 570

\bb
van den Bergh, S., 1967, AJ 72, 70

\bb
Vink, J.S., \& de Koter, A., 2002, A\&A submitted

\bb
Vink, J.S., de Koter, A., \& Lamers H.J.G.L.M., 1999, A\&A 350, 181

\bibitem[Vink et al. 2000]{vink00}
Vink, J.S., de Koter, A., \& Lamers, H.J.G.L.M, 2000, A\&A 362, 295
 
\bb
Vink, J.S., de Koter, A., \& Lamers, H.J.G.L.M, 2001, A\&A 369, 574

\bb 
Willson, L.A. \& Bowen, G.H., 1984, Nature 312, 429

\bb
Yong, H., Demarque, P. \& Yi, S., 2000, ApJ 539, 928

\bb
Zoccali M., Cassisi S., Bono G., et al. 2000, ApJ 538, 289

\end{thebibliography}
\end{document}